\documentstyle[12pt,epsfig]{article}

\newlength{\dinwidth}
\newlength{\dinmargin}
\begin{document}
\def\bold#1{\setbox0=\hbox{$#1$}%
     \kern-.025em\copy0\kern-\wd0
     \kern.05em\%\baselineskip=18ptemptcopy0\kern-\wd0
     \kern-.025em\raise.0433em\box0 }
\def\slash#1{\setbox0=\hbox{$#1$}#1\hskip-\wd0\dimen0=5pt\advance
       \dimen0 by-\ht0\advance\dimen0 by\dp0\lower0.5\dimen0\hbox
         to\wd0{\hss\sl/\/\hss}}
\newcommand{\be}{\begin{equation}}
\newcommand{\ee}{\end{equation}}
\newcommand{\bea}{\begin{eqnarray}}
\newcommand{\eea}{\end{eqnarray}}
\newcommand{\nn}{\nonumber}
\newcommand{\dd}{\displaystyle}
\newcommand{\bra}[1]{\left\langle #1 \right|}
\newcommand{\ket}[1]{\left| #1 \right\rangle}
\newcommand{\spur}[1]{\not\! #1 \,}
\thispagestyle{empty}
\vspace*{1cm}
\rightline{BARI-TH/02-429}
\rightline{January 2002}
\vspace*{2cm}
\begin{center}
  \begin{LARGE}
  \begin{bf} 
On three-body $B^0 \to D^{*-} D^{(*)0} K^+$ decays \\ 
\vspace*{0.3cm}
and couplings of heavy mesons\\
\vspace*{0.3cm}
to light pseudoscalar mesons
  \end{bf}
  \end{LARGE}
\end{center}  
\vspace*{8mm}
\begin{center}
\begin{large}
P. Colangelo and F. De Fazio
\end{large}
\end{center}  

\begin{center}
\begin{it}
 Istituto Nazionale di Fisica Nucleare, Sezione di Bari, Italy
\end{it}
\end{center}
\begin{quotation}
\vspace*{1.5cm}
\begin{center}
  \begin{bf}
  Abstract\\
  \end{bf}  
\end{center} 
\vspace*{0.5cm}
\noindent
We analyze the decay modes $B^0 \to D^{*-} D^0 K^+$ and
$B^0 \to D^{*-} D^{*0} K^+$ and, using the available experimental data, 
we find bounds for the constants $g$ and $h$ describing the strong coupling  of
heavy mesons to light pseudoscalar mesons. Both the  decay channels are 
dominated by broad $L=1$ charm resonances; the dominance is effective also
in  $B^0 \to D^{-} D^0 K^+$ and $B^0 \to D^{-} D^{*0} K^+$. 

\vspace*{0.5cm}
\end{quotation}

\newpage
\baselineskip=18pt
\vspace{2cm}
\noindent
\section{Introduction}
Recently, the BaBar Collaboration  
has observed  the three-body $B^0$ decay modes
\begin{eqnarray}
B^0 &\to& D^{*-} D^0 K^+    \label{channel1}\\
B^0 &\to& D^{*-} D^{*0} K^+ \label{channel2}
\end{eqnarray}
measuring the branching fractions \cite{Aubert:2001wz}:
\begin{eqnarray}
{\cal B}(B^0 \to D^{*-} D^0 K^+) &= & (2.8 \pm 0.7 \pm 0.5) \;\; 10^{-3}
\label{eq:res1}\\
{\cal B}(B^0 \to D^{*-} D^{*0} K^+) &= & (6.8 \pm 1.7 \pm 1.7) \;\; 10^{-3}
\;\;\; .
\label{eq:res2}
\end{eqnarray}
For the decay channel $B^0 \to D^{*-} D^0 K^+$, a measurement has
also been reported by the Belle Collaboration \cite{belle}:
\begin{equation}
{\cal B}(B^0 \to D^{*-} D^0 K^+) =  (3.2 \pm 0.8 \pm 0.7) \;\; 10^{-3}
\;\;\; . \label{eq:res1a}
\end{equation}
The results (\ref{eq:res1}), (\ref{eq:res2}) and (\ref{eq:res1a})  
(although preliminary) represent a significant improvement with respect to the 
previously available data,  obtained by  the CLEO Collaboration:
${\cal B}(B^0 \to D^{*-} D^0 K^+) =  (0.45^{+0.25}_{-0.19}\pm 0.08) \;\; 
10^{-2}$ and
${\cal B}(B^0 \to D^{*-} D^{*0} K^+) = (1.30^{+0.61}_{-0.47} \pm 0.27) \;\; 
10^{-2}$ \cite{cleoconf}. We expect that the experimental analysis  of
the processes (\ref{channel1}) and (\ref{channel2}),
including the study of the Dalitz plot,
will be further pursued in the forthcoming future. The purpose of this note
is to interpret the observations made so far, using the results  
(\ref{eq:res1})-(\ref{eq:res1a})
together with the existing datum on the two-body 
decay $B^0 \to D^{*-} D_s^{*+}$ 
\cite{Groom:in}:
\begin{equation}
{\cal B}(B^0 \to D^{*-} D_s^{*+}) =   (19 \pm 6) \;\; 10^{-3} \;\;\;. 
\label{eq:res3}
\end{equation}

The interest for $B$ transitions
into a pair of $D^{(*)}$ and a Kaon is manifold.
It has been proposed to use the modes
$B^0 ({\bar B^0}) \to D^- D^+ K_S$ and
$B^0 ({\bar B^0}) \to D^{*-} D^{*+} K_S$ (analogous to (\ref{channel2}))
to investigate CP violation effects in 
neutral $B$ decays at the B factories
\cite{Charles:1998vf}.
Such processes are induced at the quark level by  
the transitions  $b \to c \bar c s$ and $\bar b \to c \bar c \bar s$ 
and are Cabibbo-favoured
as in the case of $B^0 ({\bar B^0}) \to J/\psi K_S$, with a tiny
penguin contribution. Studies of the time-dependent 
Dalitz plot would provide us with information about the weak  mixing angles,
namely the phase $\beta$ related to the $B^0 - {\bar B^0}$ mixing.
In particular, since  amplitudes with different strong phases 
corresponding to various intermediate states contribute to the three-body
$B \to D^{(*)} D^{(*)} K_S$ decays,
one envisages the possibility of measuring both
$\sin(2 \beta)$ and $\cos(2 \beta)$  by suitable Dalitz plot analyses
\cite{Charles:1998vf}.

\begin{figure}[h]
\begin{center}
\vspace*{-1.cm}
\mbox{\epsfig{file=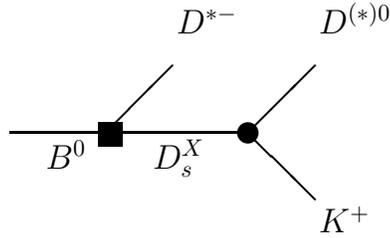, width=8.5cm}}
\vspace*{-0.5cm}
\end{center}
\caption{\baselineskip 15pt
Diagram contributing to the decay 
$B^0 \to  D^{*-}  D^{(*)0} K^+$. The box represents a weak transition,
the dot a strong vertex.}
\vspace*{1.0cm}
\label{diagrams}
\end{figure}

Another  reason of interest  concerns the possibility of carrying out
tests of factorization 
for nonleptonic $B$ decays. It is reasonable to assume that the modes
(\ref{channel1}) and (\ref{channel2})
mainly proceed through two-body intermediate states, such as
\begin{equation}
B^0 \to D^{*-} D_s^X \,\,\, , \label{eq:twob}
\end{equation}
followed by the strong transition  
\begin{equation}
D_s^X \to D^{(*)0} K^+  . \label{eq:strongtrans}
\end{equation}
$D_s^X$ are charmed strange mesons; a typical
diagram is depicted in fig.\ref{diagrams}.
In the factorization approximation the amplitude of the process in 
(\ref{eq:twob})  is expressed as the product
of the semileptonic $B^0 \to D^{*-}$ 
matrix element and the $D_s^X$ current-vacuum
matrix element.
%
In the
infinite charm quark mass limit, the only contributions
with non-vanishing  $D_s^X$ current-vacuum matrix elements 
correspond to the states $D_s^X=D_s^*$ and $D_{s0}$ 
(with their radial excitations) for 
$B^0 \to D^{*-} D^0 K^+$, and  $D_s^X=D_s^*$, 
$D_s$  and $D^*_{s1}$ (together with their radial excitations)
for $B^0 \to D^{*-} D^{*0} K^+$.
$D_{s0}$ and $D^*_{s1}$ are  positive parity mesons
belonging to the $s_\ell^P={1\over 2}^+$ heavy meson $(\bar s c)$ doublet,
$s_\ell^P$ being the spin-parity of the light degrees of freedom in the meson.
Therefore, the number of independent amplitudes contributing to
(\ref{channel1}) and (\ref{channel2}) is limited, and it is possible
to study relations, e.g., with the mode 
$B^0 \to D^{*-} D^{*+}_s$ for which  the experimental datum 
(\ref{eq:res3}) is available.
Moreover, one can exploit a possible
dominance of the positive parity intermediate
states to study the features of these (so far) unobserved states. 

There is a further reason of interest in
the processes (\ref{channel1}) and (\ref{channel2}). If the main mechanism for 
the three-body $D^{*-} D^{(*)0} K^+$
final states is the production of a pair of 
$D^{*-} D_s^X$ mesons followed by the strong transition 
(\ref{eq:strongtrans}), one can use such  decay modes 
to access the couplings of heavy mesons to light pseudoscalar states.
For these  quantities little experimental information is currently available.
The CLEO Collaboration has provided the  first determination 
of the  strong coupling constant 
$g_{D^* D \pi}$ governing  the transition $D^{*+}\to D^0 \pi^+$,
using the recent measurement of the total width of the 
$D^{*+}$ meson \cite{cleonew}
\be
\Gamma(D^{*+})=96 \pm 4 \pm 22 \;\; KeV \;\;\; \footnote{This result 
updates the upper
bound provided by the ACCMOR Collaboration: $\Gamma(D^{*+}) <131 \,\,
KeV $ at 90$\%$ c.l.\cite{accmor}.} \label{expwid}
\ee
together with   the experimental branching fraction 
${\cal B}(D^{*+} \to D^0 \pi^+)=(67.7 \pm 0.5) \; 10^{-2}$ 
\cite{Groom:in}.
The result for the coupling, defined by the matrix element
\be
<D^0(k) \pi^+(q)|D^{*+}(p,\epsilon)>=g_{D^* D \pi} \, 
\epsilon \cdot q  \label{defgddp}
\ee
($\epsilon$ is the $D^*$ polarization vector), is:
\be
g_{D^* D \pi}=17.9 \pm 0.3 \pm 1.9 \label{expgddp} \,\,.
\ee
Rewriting $g_{D^* D \pi}$ in terms of an effective coupling 
$g_D$:
\be
g_{D^*D \pi}={2 \sqrt{m_D m_{D^*}} \over f_\pi} g_D = 
{2 \sqrt{m_D m_{D^*}} \over f_\pi} g \; (1 + {\cal O}({1\over m_c}) )\,\, , 
\label{defg}
\ee
one translates the result (\ref{expgddp}) into
\be
g_D=0.59 \pm 0.01 \pm 0.07 \label{gexp} \,.
\ee
In the heavy quark limit the parameter $g$ in (\ref{defg})
describes  the strong
coupling of charmed mesons as well as of beauty mesons to the members
of the octet of light pseudoscalars; therefore, neglecting $SU(3)_F$
breaking effects, this parameter enters 
some matrix elements governing the transitions in (\ref{eq:strongtrans}).
In addition, together with analogous couplings,  
$g$ represents a basic quantity  in the heavy-quark chiral effective theory
\cite{hqet_chir,Casalbuoni:1996pg},
and therefore it is worth searching information about
it from all available  experimental data, and comparing the results with the
predictions that vary in the rather wide range $0.2<g<0.7$ \cite{khod}.
This is a purpose of the present note.

In the next section we  analyze the decay modes
(\ref{channel1}) and (\ref{channel2}) and discuss how to
access the relevant strong couplings. Numerical results follow in section 3. 
The conclusions are drawn at the end of the note.

\section{Decay modes $B^0 \to D^{*-} D^0  K^+$ 
and $B^0 \to D^{*-} D^{*0} K^+$}

Let us consider the processes (\ref{channel1}) and (\ref {channel2}):
\bea
B^0(p) &\to&  D^{*-}(p_-, \epsilon_-) D^0(p_D)  K^+(q) \nonumber \\
B^0(p) &\to&  D^{*-}(p_-, \epsilon_-) D^{*0}(p_{D^*}, \epsilon) K^+(q) 
\nonumber
\label{modes}
\eea
with the momenta
$p=m_B v$, $p_-=m_{D^*} v_-$ and $p_{D^{(*)}}=m_{D^{(*)}} w$ expressed  
in terms of the heavy-meson four-velocities $v,\;\; v_-$ and $w$. 
Neglecting penguin contributions, the processes are 
governed by the effective weak Hamiltonian:
\be
H_W={G_F \over \sqrt{2}}V_{cs} V_{cb}^* \, a_1 \,{\bar b}
\gamma_\mu (1-\gamma_5) c \, {\bar c} \gamma^\mu (1-\gamma_5) s 
\label{hamilt}
\ee
where $G_F$ is the Fermi constant, $V_{ij}$ are CKM matrix
elements and the parameter $a_1$ reads
$a_1=\left(c_1+ \displaystyle{c_2 \over N_c} \right)$,
with $c_{1,2}$ 
short-distance Wilson coefficients and $N_c$ the number of colors.
Dalitz plot variables of the decays can be defined:
\bea
s &=& (p_{D^{(*)}}+q)^2 \nonumber \\
s_- &=& (p_-+q)^2 \label{dalitz} 
\eea
and a set of invariant variables, in terms of the four-velocities 
$v,v_-$ and $w$, can be introduced:
\bea
v \cdot v_- &=& { m_B^2+m_{D^*}^2-s \over 2 m_B m_{D^*}} \nonumber \\
v \cdot w  &=& {m_B^2+m^2_{D^{(*)}}-s \over 2 m_B m_{D^{(*)}} }
\nonumber \\
v_- \cdot w &=& {m_B^2+m_K^2-s-s_- \over 2 m_{D^*} m_{D^{(*)}} }
\label{dotprod} \\
v \cdot q &=& { s + s_- -m_{D^*}^2-m_{D^{(*)}}^2 \over 2 m_B} 
\nonumber \\
v_- \cdot q &=& {s_- -m_{D^*}^2-m_K^2 \over 2 m_{D^*}}
\nonumber \\
w \cdot q &=& {s -m_K^2 -m_{D^{(*)}} \over 2 m_{D^{(*)}} }
\,\,\;\;\;. \nonumber 
\eea
In the plane $(s, s_-)$ the accessible kinematical region is defined by
the conditions 
\bea
(m_{D^{(*)}}+m_K)^2 \le & s & \le (m_B -m_{D^*})^2 \label{intlim} \nonumber \\
(s_-)_- \le &s_-& \le (s_-)_+ 
\eea
where
\bea
(s_-)_\pm =m_{D^*}+m_K^2&-&{1 \over 2 s}
\Big[ (s-m_B^2+m_{D^*}^2)(s+m_K^2-m_{D^{(*)}}^2) \nonumber \\
&& \mp \lambda^{1/2}(s, m_K^2, m_{D^{(*)}}^2) \lambda^{1/2}(s, m_B^2,
m_{D^*}^2) \Big] \,\,,
\label{smlim} 
\eea
$\lambda$ being the triangular function.

We assume that the decays $B^0 \to  D^{*-} D^{(*)0} K^+$  proceed through
polar diagrams such as the one depicted in fig.\ref{diagrams}, computed
adopting the factorization approximation for the weak transition. 
In the case of $B^0 \to   D^{*-} D^{0} K^+$, the pole can be either a vector
($J^P=1^-$) meson: $D_s^{*+}$, or a scalar ($J^P=0^+$)  meson: $D_{s0}^+$,
with their radial excitations.
For the decay $B^0 \to  D^{*-} D^{*0} K^+$, the
possible poles are: $D_s^{*+}$ ($J^P=1^-$), $D_s^+$ ($J^P=0^-$) and
$D_{1s}^{*+}$ ($J^P=1^+$) and their radial excitations. 
Therefore, the calculation of the amplitudes in
fig.\ref{diagrams} requires the  strong vertices
\bea
<D^0(p_D) K^+(q)|D_s^* ( p_{D^*_s},\epsilon_s)> &=& g_{D^*_s D K}
(\epsilon_s \cdot q) \nonumber \\
<D^0(p_D) K^+(q)|D_{s0}^+(p_{D_{s0}})> &=& g_{D_{s0} D K}
\label{strongcoupl} \nonumber \\
<D^{*0}(p_{D^*}, \epsilon) K^+(q)|D_s^*(p_{D^*_s},\epsilon_s)>
&=& i \,
{g_{D^*_s D^* K} \over m_{D^*_s}}  
\, \epsilon_{\tau \theta \phi \psi} \epsilon_s^\tau
\epsilon^{* \theta} p_{D^*_s}^\phi q^\psi \\
<D^{*0}(p_D^*, \epsilon) K^+(q)|D_s(p_{D_s})> &=&
g_{D^* D_s K} (\epsilon^* \cdot q) \nonumber \\
<D^{*0}(p_D^*, \epsilon) K^+(q)|D^{*+}_{s1}(p_{D^*_{s1}},\epsilon_s)> 
&=& {g_{D^*_{s1} D^* K} \over m_{D^*_{s1}}} \; (\epsilon^* \cdot \epsilon_s)
(p_{D^*} \cdot q) 
\,\, ,\nonumber 
\eea
and analogous matrix elements involving radial $D_s^X$ resonances.
In the heavy quark limit, all the couplings in (\ref{strongcoupl}) 
can be expressed in
terms of two different coupling constants $g$ and $h$, for negative and 
positive parity $D_s^X$ states, respectively.
This can be shown considering the effective lagrangian
describing the interactions of heavy mesons with the light pseudoscalars. 
In the limit $m_Q \to \infty$ the heavy quark in the heavy mesons only
acts as a static colour source, and the 
gluons decouple from the heavy quark spin $s_Q$, thus implying a
$SU(2 N_f)$ spin-flavour  symmetry \cite{hqet,DeFazio:2000up}.
At the opposite energy scale, for vanishing masses of the up, down and
strange quarks, the QCD $SU(3)_L \times SU(3)_R$ chiral
symmetry is spontaneously broken, the Goldstone bosons being 
the octet of the light pseudoscalar mesons.
Both the heavy quark spin-flavour and the chiral symmetries can be
realized in a  QCD effective lagrangian \cite{hqet_chir}, where
the term describing the strong interactions of the heavy negative and 
positive parity mesons with the light pseudoscalars reads:
\begin{equation}
{\cal L}_I = i \;
g \; Tr\{H_b \gamma_\mu \gamma_5 {\cal A}^\mu_{ba} {\bar H}_a\} 
+  \;[ \; i \;
h \; Tr\{H_b \gamma_\mu \gamma_5 {\cal A}^\mu_{ba} {\bar S}_a\}
+ \; h.c. \;] \;\; .  \label{L}
\end{equation}
The fields $H_a$ in (\ref{L})
describe the negative parity  $J^P=(0^-,1^-)$ ${\bar q }Q$ meson doublet,
with $s_\ell^P= {1\over 2}^-$:
\begin{equation}
H_a = \frac{(1+{\rlap{v}/})}{2}[P_{a\mu}^*\gamma^\mu-P_a\gamma_5] \;\;,
\label{neg}
\end{equation}
the operators $P^{*\mu}_a$ and $P_a$ respectively annihilating the 
$1^-$  and $0^-$  mesons of four-velocity $v$ ($a=u,d,s$ is a light flavour 
index). Analogously, the  fields $S_a$ describe the positive parity  states,
with $s_\ell^P= {1\over 2}^+$:
\begin{equation}
S_a = \frac{(1+{\rlap{v}/})}{2}[P_{a\mu}^{\prime *}\gamma^\mu \gamma_5 - 
P^\prime_a] \;\;.
\label{pos}
\end{equation}
The octet of the light pseudoscalar mesons is included in (\ref{L}) through
the field $\displaystyle \xi=e^{i {\cal M} \over f_\pi}$, with
\begin{equation}
{\cal M}=
\left (\begin{array}{ccc}
\sqrt{\frac{1}{2}}\pi^0+\sqrt{\frac{1}{6}}\eta & \pi^+ & K^+\nonumber\\
\pi^- & -\sqrt{\frac{1}{2}}\pi^0+\sqrt{\frac{1}{6}}\eta & K^0\\
K^- & {\bar K}^0 &-\sqrt{\frac{2}{3}}\eta
\end{array}\right )
\end{equation}
and $f_{\pi}=131 \; MeV$.
Finally, the operator $\cal A$ in (\ref{L}) reads
\begin{equation}
{\cal A}_{\mu ba}=\frac{1}{2}\left(\xi^\dagger\partial_\mu \xi-\xi
\partial_\mu \xi^\dagger\right)_{ba} \; .
\end{equation}
\par

From  the definitions in (\ref{strongcoupl}) and from 
eq.(\ref{L}) it is straightforward to derive the relations
\bea
g_{D^*_s D K} &=& {2 \sqrt{m_{D^*_s} m_D} \over f_K}\, g \nonumber \\
g_{D_{s0} D K} &=& -\sqrt{m_{D_{s0}}m_D} \, {m_{D_{s0}}^2-m_D^2 \over
m_{D_{s0}} } \, {h \over f_K} \nonumber \\
g_{D^*_s D^* K} &=& {2 m_{D^*_s} \over f_K}\, g \label{gh} \\
g_{D^* D_s K} &=& {2 \sqrt{m_{D^*} m_{D_s}} \over f_K} \, g \nonumber \\
g_{D^*_{s1} D^* K} &=& -{2 \sqrt{m_{D^*_{s1}} m_{D^*} } \over f_K}\, h
\,\, , \nonumber 
\eea
where we have kept some $SU(3)$ flavor breaking terms in the
masses of the $D_s^X$ mesons and in the leptonic constant $f_K$.

On the other hand, in the factorization approximation
the calculation of the weak transition (\ref{eq:twob}) requires 
the semileptonic $B^0 \to D^{*-}$ matrix element 
and the decays constant of the poles
$D_s^X$. In the heavy quark limit, the former is given
in terms of the Isgur-Wise function  $\xi$:
\begin{eqnarray}
<D^{*-}(v_-,\epsilon_-)|V^\mu-A^\mu|B(v)>&=&
\sqrt{m_B m_{D^*}} \; \xi(v \cdot v_-) \; \epsilon^*_{-\alpha} \; 
\nonumber \\
&\big(&- i \varepsilon ^{\rho \alpha \lambda \mu} v_\rho v_{-\lambda}
- (1+v \cdot v_-) g^{\alpha \mu}+ v^\alpha v^\mu_- \big) 
\end{eqnarray}
while the decay constants are defined by 
\bea
<D_s^+(p_{D_s})|{\bar c} \gamma^\mu(1-\gamma_5)s|0>&=&i
f_{D_s}p_{D_s}^\mu  \nonumber \\
<D_s^{*+}(p_{D^*_s},\epsilon_s)|{\bar c} \gamma^\mu(1-\gamma_5)
s|0>&=&f_{D^*_s} m_{D^*_s} \epsilon_s^{* \mu} \nonumber \\
<D_{s0}^+(p_{D_{s0}})|{\bar c} \gamma^\mu(1-\gamma_5)s|0>&=&i
f_{D_{s0}}p_{D_{s0}}^\mu \\
<D_{s1}^{*+}(p_{D^*_{s1}},\epsilon_s)|{\bar c} \gamma^\mu(1-\gamma_5)
s|0>&=&f_{D^*_{s1}} m_{D^*_{s1}} \epsilon_s^{* \mu} \,\, . \nonumber 
\label{decayconst}
\eea
In the heavy quark limit, the leptonic constants
$f_{D_s}$ and $f_{D^*_s}$, as well as $f_{D_{s0}}$ and $f_{D^*_{s1}}$,
are simply related. 

It is now straightforward to work out  the amplitude of 
$B^0 \to D^{*-} D^0 K^+$ proceeding  {\it via} the $D_s^*$ intermediate
state:
\bea
{\cal A}_1 = {i {\cal K} \, f_{D^*_s} m_{D^*_s}\, g_{D^*_s D K} \over s-
m_{D^*_s}^2+i m_{D^*_s} \Gamma_{D^*_s}} \, 
\xi \left(v \cdot v_-\right)\, \epsilon^{* \nu}_-  
\label{a1} \\
\left( -q^\mu +
{(m_B v-m_{D^*} v_-)\cdot q \over m_{D^*_s}} \, {(m_B
v-m_{D^*} v_-)^\mu \over m_{D^*_s}} \right) \nonumber \\
\left\{ i \, \epsilon_{\mu \nu
\alpha \beta} v_-^\alpha v^\beta -g_{\mu \nu} (1+ v \cdot v_-) +v_\nu
(v_-)_\mu \right\} \nonumber
\eea
with
${\cal K}=\displaystyle{G_F \over \sqrt{2}}V_{cs}V_{cb}^* a_1
\sqrt{m_B m_{D^*}}$ and $\Gamma_{D^*_s}$ the $D^*_s$ decay width.
Analogously, the amplitude  ${\cal A}_2$ relative to the  $D_{s0}$
contribution to  $B^0 \to D^{*-} D^0 K^+$ reads:
\bea
{\cal A}_2&=&-{{\cal K} \, f_{D_{s0}} m_{D_{s0}} \, g_{D_{s0}D K} \over s
-m_{D_{s0}}^2+i m_{D_{s0}} \Gamma_{D_{s0}}} \,
\xi \left(v \cdot v_-\right)\, \epsilon^{* \nu}_- \label{a2} \\
&&\left\{-{(m_B v-m_{D^*} v_-)_\nu \over m_{D_{s0}} } (1 +v \cdot v_-)
+v_\nu {(m_B v-m_{D^*} v_-)\cdot v_- \over m_{D_{s0}} } \right\} \,.
\nonumber 
\eea
As for  $B^0 \to D^{*-} D^{*0} K^+$, the amplitudes
${\cal A}^*_1, {\cal A}^*_2$ and ${\cal A}^*_3$ corresponding to the $D_s^*$,
$D_s$ and $D^*_{s1}$ intermediate states are given by:
\bea
{\cal A}_1^*&=&
 { {\cal K} \, f_{D^*_s} m_{D^*_s}\, g_{D^*_s D^* K} \over s-
m_{D^*_s}^2+i m_{D^*_s} \Gamma_{D^*_s}} \,
\xi \left(v \cdot v_-\right)\, \epsilon^{* \nu}_-  
\epsilon^\mu_{\theta \phi \psi} \epsilon^{* \theta}
{m_{D^*}\over m_{D^*_s}}w^\phi q^\psi \nonumber \\
&&\left\{ i \, \epsilon_{\mu\nu\alpha \beta} 
v_-^\alpha v^\beta -g_{\mu \nu} (1+ v \cdot v_-) +v_\nu
(v_-)_\mu \right\}\,; \\ \label{a1star}  
{\cal A}_2^* &=&
{{\cal K}f_{D_s} g_{D^* D_s K} \over s -m_{D_s}^2 +i m_{D_s} \Gamma_{D_s}}
\, \xi \left(v \cdot v_-\right)\, (m_B + m_{D^*}) \, (\epsilon \cdot q) \,
(\epsilon^*_- \cdot v) \,; \\ \label{a2star}
{\cal A}_3^*&=&
i {{\cal K} f_{D^*_{s1}} m_{D^*} g_{D^*_{s1} D^* K} \over s
-m_{D^*_{s1}}^2 +i m_{D^*_{s1}} \Gamma_{D^*_{s1}} }\,
\, \xi \left(v \cdot v_-\right)\, (w \cdot q) \, \epsilon_-^{*\nu}
\epsilon^{*\tau} \nonumber  \\
&&\big\{-i \epsilon_{\tau \nu \alpha \beta}v_-^\alpha v^\beta + g_{\nu
\tau} (1+ v \cdot v_-)-v_\nu (v_-)_\tau \nonumber \\
&-&(1 + v \cdot v_-) {(m_B v-m_{D^*} v_-)_\nu \over m_{D^*_{s1}} }
{(m_B v-m_{D^*} v_-)_\tau \over m_{D^*_{s1}} } \nonumber \\
&+&{(m_B v-m_{D^*} v_-) \cdot v_- \over m_{D^*_{s1}} } 
{(m_B v-m_{D^*} v_-)_\tau \over m_{D^*_{s1}} } v_\nu \big\}
\,. \label{a3star}
\eea

Expressions analogous to eqs.(\ref{a1})-(\ref{a3star}), with appropriate 
masses, widths, leptonic constants and strong couplings, hold for the
contributions of the radial excitations of negative and positive parity mesons.
Such contributions are suppressed by the small
numerical values of the leptonic constants and of the effective couplings. 
This can be shown, for example, using the relativistic constituent quark model 
in ref.\cite{Colangelo:1990rv}, where
one obtains: $f_{D^\prime_s}/f_{D_s}\simeq 0.73$,
$D_s^\prime$ being the first radial excitation of $D_s$.
In the same model, using the method described in \cite{noipotmod}, 
one obtains: $g_{D^\prime_s D^* K}/g_{D_s D^* K}\simeq 0.32$. Analogous
reductions occur for positive parity states. A further suppression
is due to the large decay width of the excited states. Therefore,
one can conclude that the first radial excitations contribute 
to the amplitudes of the processes (\ref{channel1}) and (\ref{channel2})
by less than $15\%$ with respect
to the contribution of the corresponding low-lying states, an uncertainty 
that can be included in the error affecting the effective couplings 
we are studying in this note.

\section{Numerical analysis and discussion}

On the basis of the above considerations, we write down the widths 
of the decay modes $B^0 \to D^{*-}  D^0  K^+$  and
$B^0 \to D^{*-}  D^{*0}  K^+$ as follows:
\be
\Gamma(B^0 \to D^{*-} D^{(*)0}  K^+)=
\int_{(m_{D^{(*)}}+m_K)^2}^{(m_B-m_{D^*})^2} d s \, 
\int_{(s_-)_-}^{(s_-)_+} ds_- \, {d \Gamma \over ds \, ds_-} \label{width}
\,\,\, , 
\ee
with
$\displaystyle
{d \Gamma \over ds \, ds_-}(B^0 \to D^{*-} D^{(*)0}  K^+)= {1\over (2 \pi)^3}
 {1\over 32 m_B^3} {\overline{|{\cal A}|^2}}$
and
\bea
{\cal A}(B^0 \to D^{*-} D^0 K^+)&=&\sum_{i=1,2} {\cal A}_i  \,\,\,\nonumber \\
{\cal A}(B^0 \to D^{*-} D^{*0} K^+)&=&\sum_{i=1,2,3} {\cal A}_i^* \,\, .
\label{ampl}
\eea 
The decay widths
 depend on the effective couplings  $g$ and $h$. They also depend on
SM parameters, such as ${G_F \over \sqrt 2} V_{cs} V^*_{cb}$, on the
leptonic constants $f_{D_s}, \dots $, on the Wilson coefficients $c_{1,2}$ 
as well as on the Isgur-Wise 
form factor $\xi$. All such parameters appear in the same combination in 
the factorized  amplitude of the two-body decay 
in (\ref{eq:res3}):
\begin{equation}
{\cal A}(B^0 \to D^{*-} D^*_s)=  {\cal K} f_{D^*_s} m_{D^*_s} 
\xi \left(v \cdot v_-\right)\, \epsilon^{* \alpha} \epsilon^{* \mu}_- 
[ i \, \epsilon_{\rho\alpha \lambda \mu} v_-^\lambda v^\rho 
+ g_{\alpha \mu} (1+ v \cdot v_-) - v_\alpha v_\nu] \,\,\, .\nonumber
\end{equation}
Therefore, in the ratios
\bea
R_D&=&{{\cal B}(B^0 \to D^{*-}  D^0 K^+) \over{\cal B}(B^0 \to
D^{*-} D_s^{*+}) } \nonumber \\ 
R_{D^*}&=&{{\cal B}(B^0 \to D^{*-} D^{*0} K^+) \over{\cal B}(B^0 \to
 D^{*-} D_s^{*+} )} 
\label{ratios}
\eea
one gets rid of the  dependences on $V_{cb}$, $V_{cs}$ and $a_1$.
As for the Isgur-Wise function, the linear form
\be
\xi(v \cdot v_-)=1-{\hat \rho}^2\,(v\cdot v_- -1) \label{xi}
\ee
is well suited due to the narrow range of momentum transfer involved 
in the decays we are considering.
A strong correlation has been observed between the measured values of 
$V_{cb}$ and the slope parameter ${\hat \rho}^2$ in the analyses
of the semileptonic $B^0 \to D^{*-} \ell \nu$ decay spectrum
and in the studies of two-body $B$ transitions in the factorization 
approximation \cite{lepwg,rosner}.
However, in the ratios (\ref{ratios}) 
such a correlation is essentially removed, and similar results are obtained
varying $\hat \rho^2$ in the range $\hat \rho^2=1.38-1.54$.

Concerning the leptonic constants, we use 
$\displaystyle{f_{D_s} \over f_{D^*_s}}=1$ and
$\displaystyle{{f_{D_{s0}} \over f_{D^*_s}}={f_{D_{s1}^*} \over
f_{D^*_s}}}=1$. The former ratio exactly holds in the infinite
charm quark limit. The latter one allow us to reduce the number
of input parameters, since a deviation from unity can be reabsorbed 
in the numerical result for the parameter $h$.
For the masses of the excited charm mesons, we use 
$m_{D_{s0}}=m_{D^*_{s1}}=m_{D_s}+\Delta$, with
 $\Delta=0.5 \,GeV$  \cite{noidelta}. 

The final set of input quantities involves the decay widths of the 
intermediate states. One can neglect the $D_s$ width
($\Gamma(D_s)=1.33\pm 0.03\, 10^{-9}\,MeV$  \cite{Groom:in}). As for $D^*_s$,
as well as for the positive parity charmed states, the widths depend on 
the effective couplings $g$ and $h$. 
Using the experimental branching fractions
${\cal B}(D^{*+}_s \to  D_s^+ \pi^0)$ and ${\cal B}(D^{*+}_s \to  D_s^+
\gamma)$, together with the central value of $g$ in (\ref{gexp}), one obtains
$\Gamma(D_s^*)=1.03 \, MeV$; consequently, we use the expression
$\Gamma(D_s^*)=1.03 \, ({g \over 0.59})^2 \, MeV$ 
in the analysis for constraining
the strong coupling $g$. Moreover,  assuming that the decay widths of
the positive parity states are saturated by two-body transitions, one gets
$\Gamma(D_{s0})=180 \, ({h\over 0.56})^2 \, MeV$ and
$\Gamma(D_{s1}^*)=165 \, ({h\over 0.56})^2 \, MeV$ \cite{noidelta}.

\begin{figure}[ht]
\begin{center}
\mbox{\epsfig{file=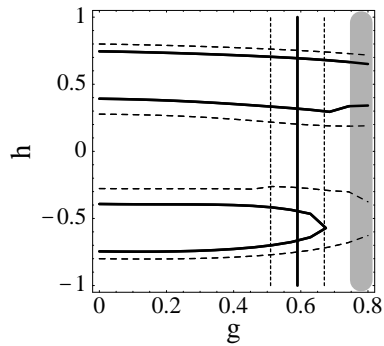, width=5cm}}
\end{center}
\caption{\baselineskip 15pt
$1-\sigma$ (continuous lines) and $2-\sigma$ (dashed lines) region in the 
$(g,h)$ plane, as 
obtained from the ratios $R_D$ and $R_{D^*}$ in (\ref{ratios}). 
The vertical lines represent the result (\ref{gexp}); the shaded
area corresponds to the region excluded by the
upper bound $g<0.76$ from ref.\cite{accmor}.}
\label{fig:constraints}
\vspace*{1.0cm}
\end{figure}

\begin{figure}[h]
\begin{picture}(100,2)(0,0)
\put(60,-40){$\displaystyle{\Large{d \Gamma \over ds \, ds_-}}$}
\end{picture}
\vskip -1.5cm
\begin{center}
\mbox{\epsfig{file=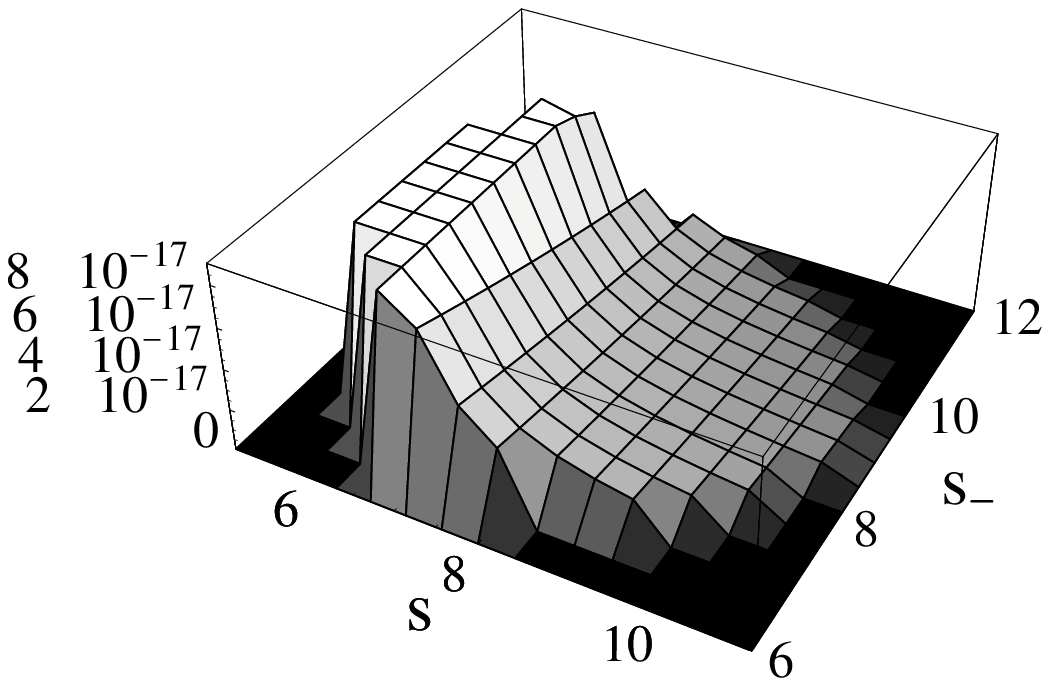, width=6.cm}}\hspace*{1.5cm}
\mbox{\epsfig{file=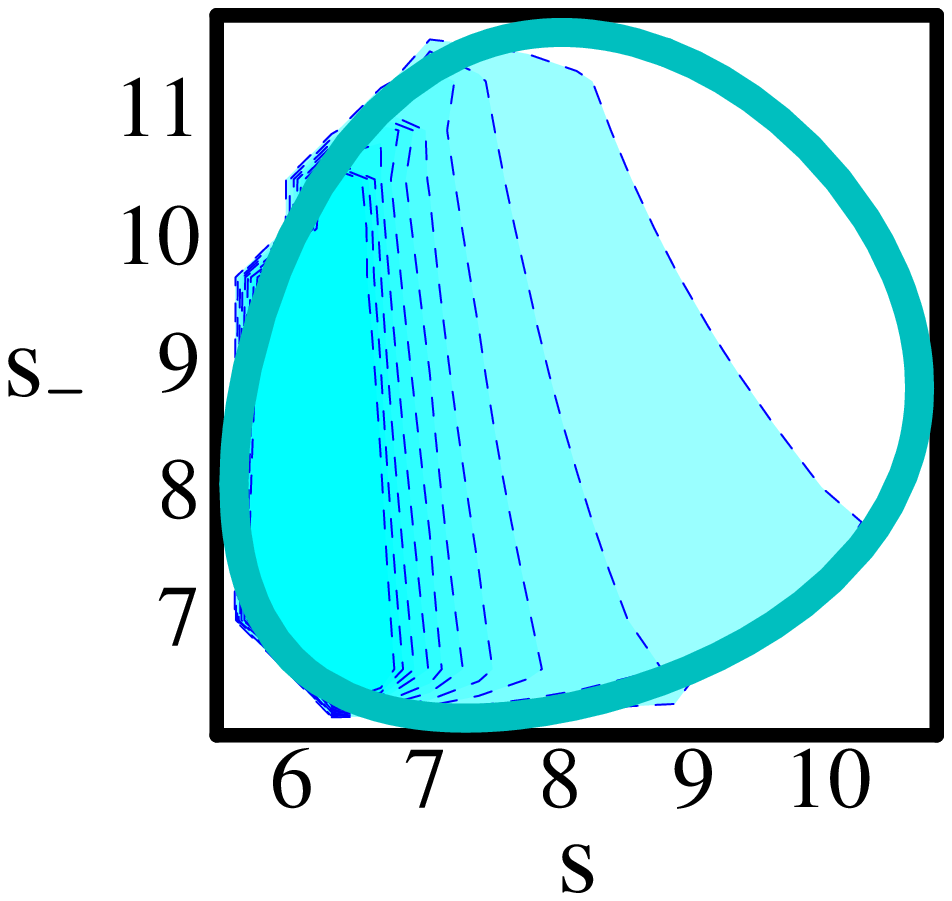, width=4cm}}
\end{center}
\caption{\baselineskip 15pt 
Differential decay width $\displaystyle {d \Gamma \over d s \, d s_-}$ (left) and
Dalitz plot (right) of the decay 
$B^0 \to D^{*-} D^0 K^+$. Units of $s$ and $s_-$ are GeV$^2$.}
\label{fig:dalitz1}
\vspace*{1.0cm}
\end{figure}

Solutions of  the equations
\bea
R_D = R_D|_{exp} &=& 0.15 \pm 0.07 \nonumber \\
R_{D^*} = R_{D^*}|_{exp} &=& 0.36 \pm 0.17 \label{eq:cond1}
\eea
in the variables $(g,h)$ are found, considering
the central values in (\ref{eq:cond1}), 
for  $(g,h)=(0.05,-0.59)$ and  $(g,h)=(0.0,+0.60)$.
The solutions for $g$ are  smaller than 
the result in (\ref{gexp}), while the results for $h$ are 
compatible with the theoretical expectations 
$h=-0.52\pm0.17$ and $h=-0.56\pm0.28$ \cite{noidelta}. 
However, before drawing  
conclusions from these results, it is worth analyzing the $1-$ and
$2-\sigma$ regions in the plane $(g,h)$, obtained considering 
the experimental errors in (\ref{eq:cond1}).
Such regions are depicted in fig.\ref{fig:constraints}. 
They are rather tightly bounded along
the $h$ direction, while the dependence on $g$ is mild and the range 
of the allowed values of $g$ extends over all the values between
$g=0$ and the CLEO result eq.(\ref{gexp}).  
Along the $h$ axis, the allowed regions correspond to $|h|=0.6 \pm 0.2$.
The conclusion is that the main contributions
to the processes (\ref{channel1}) and (\ref{channel2}) 
are related not to the $0^-$ and $1^-$, $D_s$ and $D^*_s$ intermediate states, 
but to the positive parity $0^+$ and $1^+$ states $D_{s0}$ and 
$D_{s1}^*$, since the amplitudes display a minor sensitivity to the coupling
of the negative parity intermediate states.
%
%
This is interesting from the 
phenomenological point of view, since it implies that three-body 
$B^0 \to D^{*-} D^{0} K^+$ and $B^0 \to D^{*-} D^{*0} K^+$
decay modes are well suited for separately studying the properties of the 
(so far unobserved)  $D_{s0}$  and $D_{s1}^*$ resonances
\cite{Charles:1998vf}.
The analysis can be done  by studying the  Dalitz plot 
of the three-body decay. For the mode
$B^0 \to D^{*-} D^0 K^+$ the expected differential decay width 
is depicted  in fig.\ref{fig:dalitz1}.
It has been obtained for $g=0.5$, $h=-0.6$, $a_1=1.1$, together with
$V_{cb}=0.04$ and $V_{cs}=0.974$ \cite{Groom:in}. As for $f_{D_s}$,
we use the value $f_{D_s}=240$ MeV obtained from the fit of 
(\ref{eq:res3}); it is compatible, within the errors,
with the value reported by
\cite{Groom:in}: $f_{D_s}=280 \pm 19 \pm 28 \pm 34$ MeV. The  
distribution for  $g=0.3$ is completely similar.

\begin{figure}[ht]
\begin{picture}(100,2)(0,0)
\put(60,-50){$\displaystyle{\Large{d \Gamma \over ds \, ds_-}}$}
\end{picture}
\vskip -2.5cm
\begin{center}
\mbox{\epsfig{file=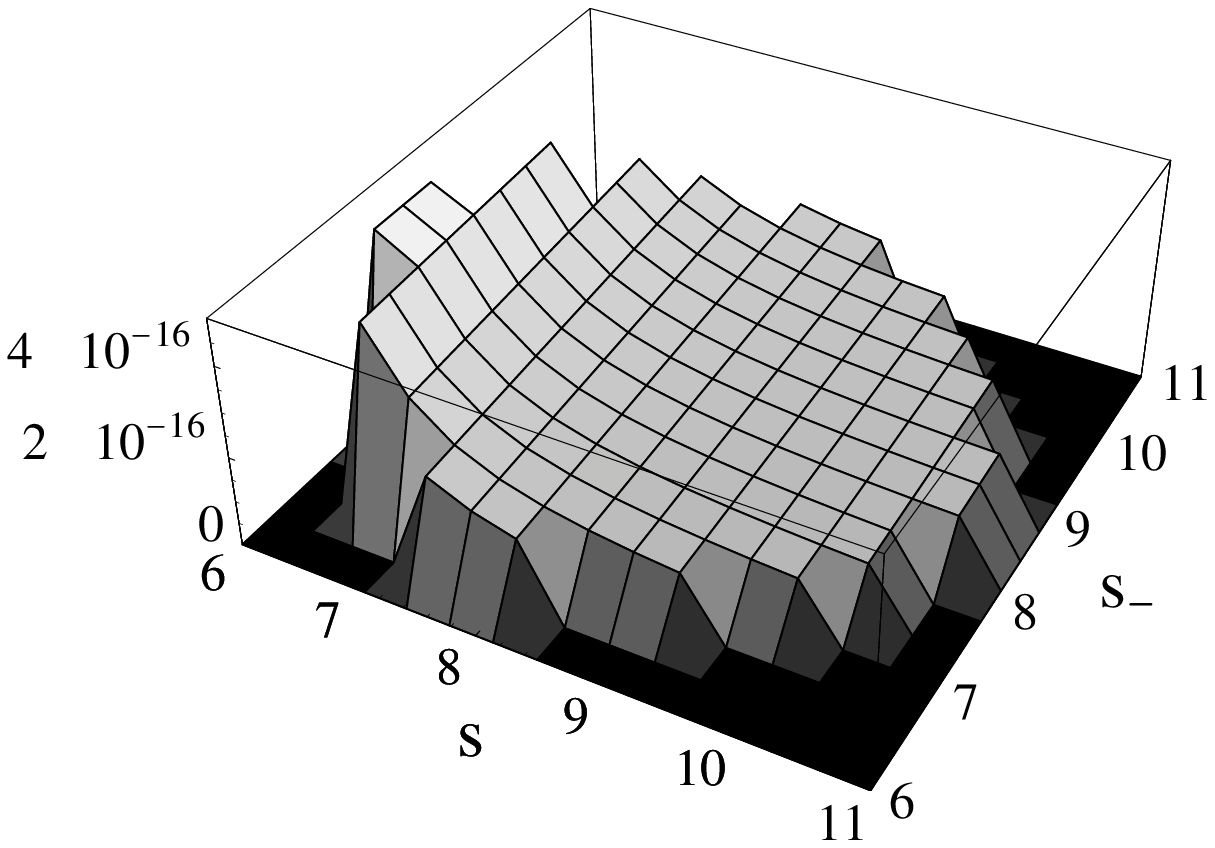, width=6cm}}\hspace*{1.5cm}
\mbox{\epsfig{file=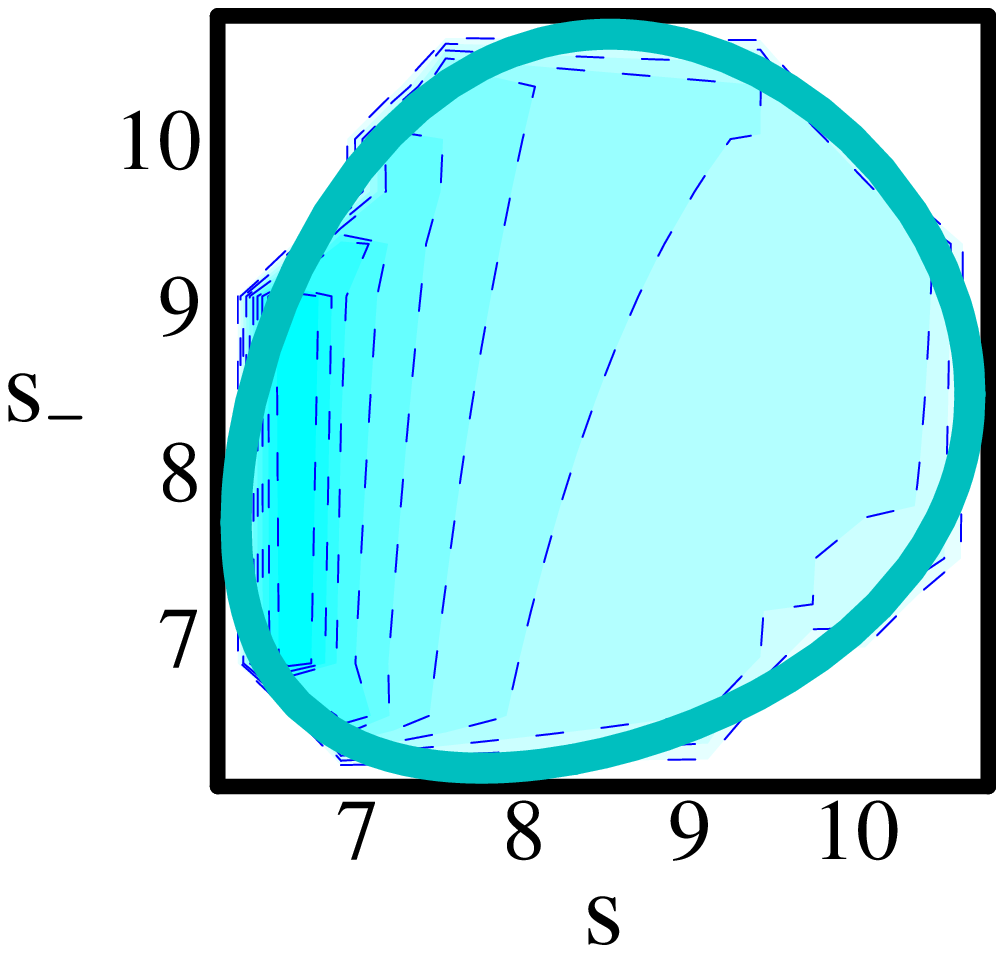, width=4cm}}
\end{center}
\caption{\baselineskip 15pt 
Differential decay width (left) and Dalitz plot (right) of the decay 
$B^0 \to D^{*-} D^{*0} K^+$. Units are as in figure \ref{fig:dalitz1}. }
\label{fig:dalitz2}
\vspace*{1.0cm}
\end{figure}

One notices that the main variation of the differential decay distribution 
occurs along the direction of the invariant $D^0 K^+$ mass,
a feature related to
the unique topology of the (Cabibbo and color allowed)
amplitudes governing the mode (\ref{channel1}).
The  Dalitz plot for 
$B^0 \to D^{*-} D^{*0} K^+$, depicted in fig.\ref{fig:dalitz2}, shows
similar features.
%
%
%

\begin{figure}[ht]
\begin{picture}(100,2)(0,0)
\put(60,-40){$\displaystyle{\Large{d \Gamma \over ds \, ds_-}}$}
\put(60,-180){$\displaystyle{\Large{d \Gamma \over ds \, ds_-}}$}
\end{picture}
\vskip -2.5cm
\begin{center}
\mbox{\epsfig{file=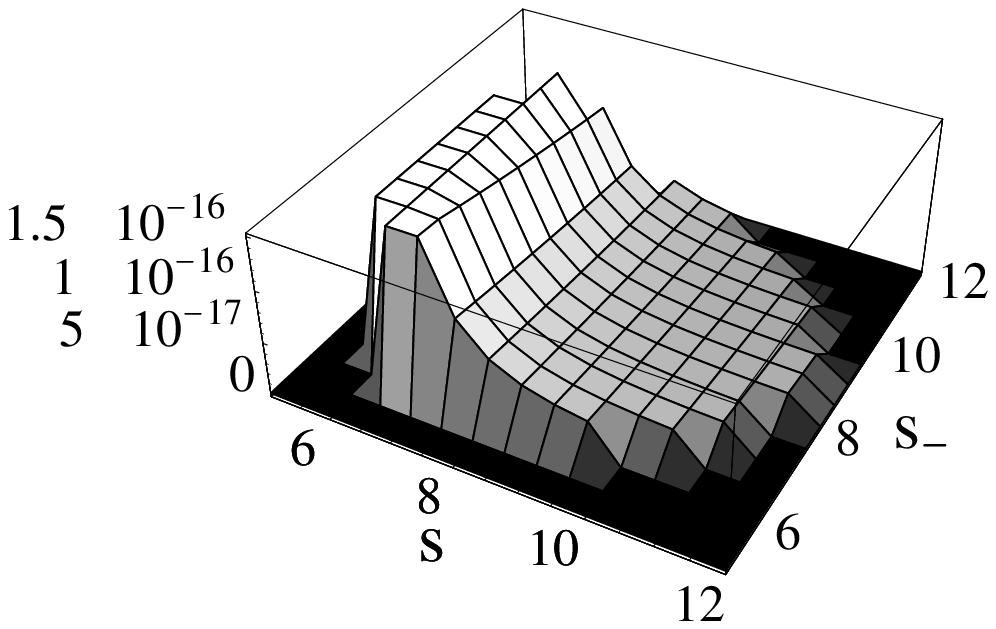, width=6cm}}\hspace*{1.5cm}
\mbox{\epsfig{file=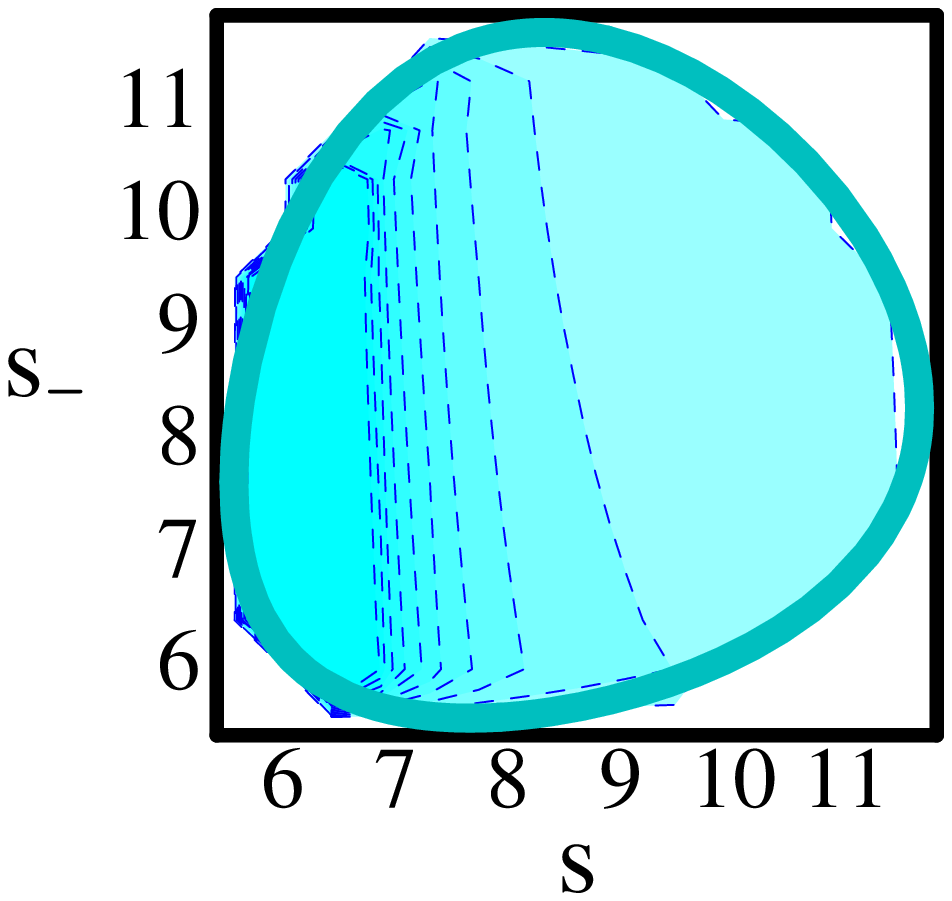, width=4cm}} \\
\mbox{\epsfig{file=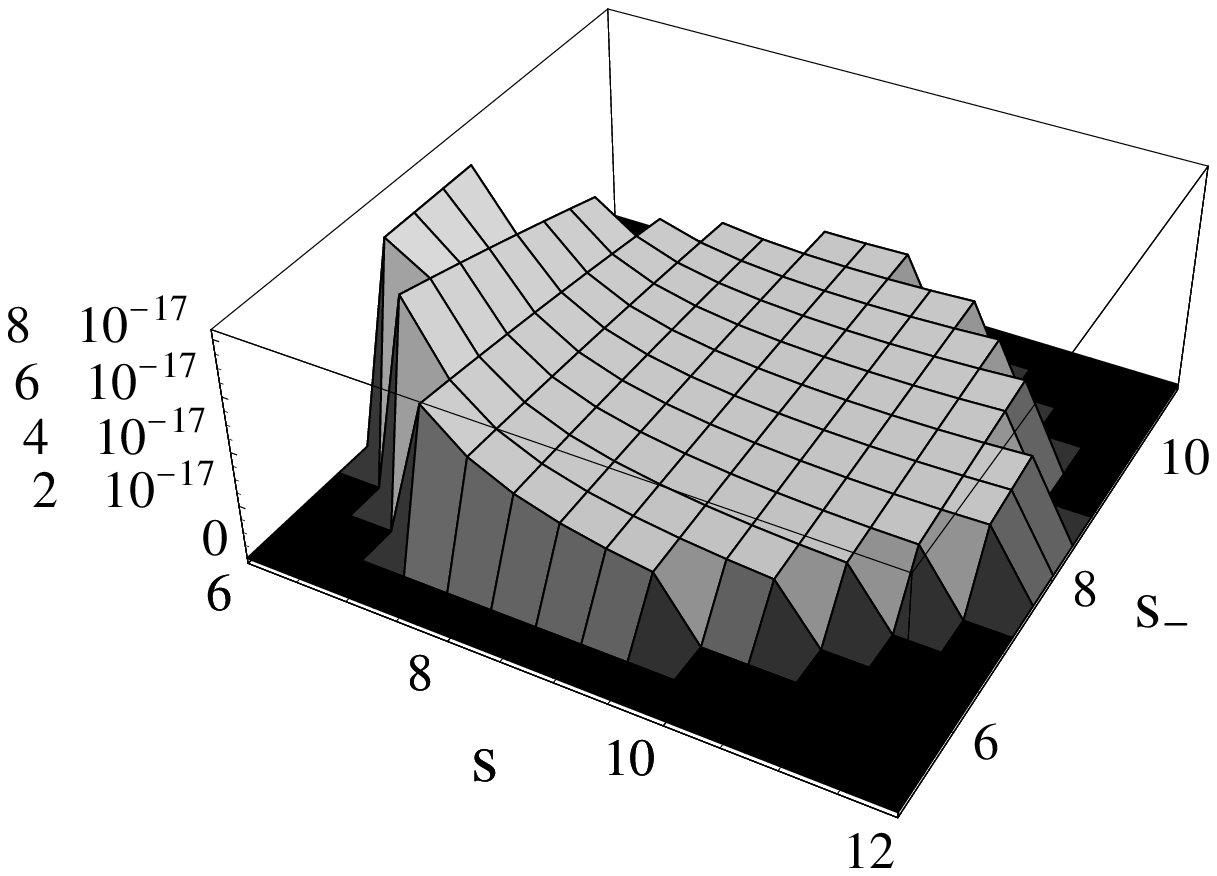,
width=6cm}}\hspace*{1.5cm}
\mbox{\epsfig{file=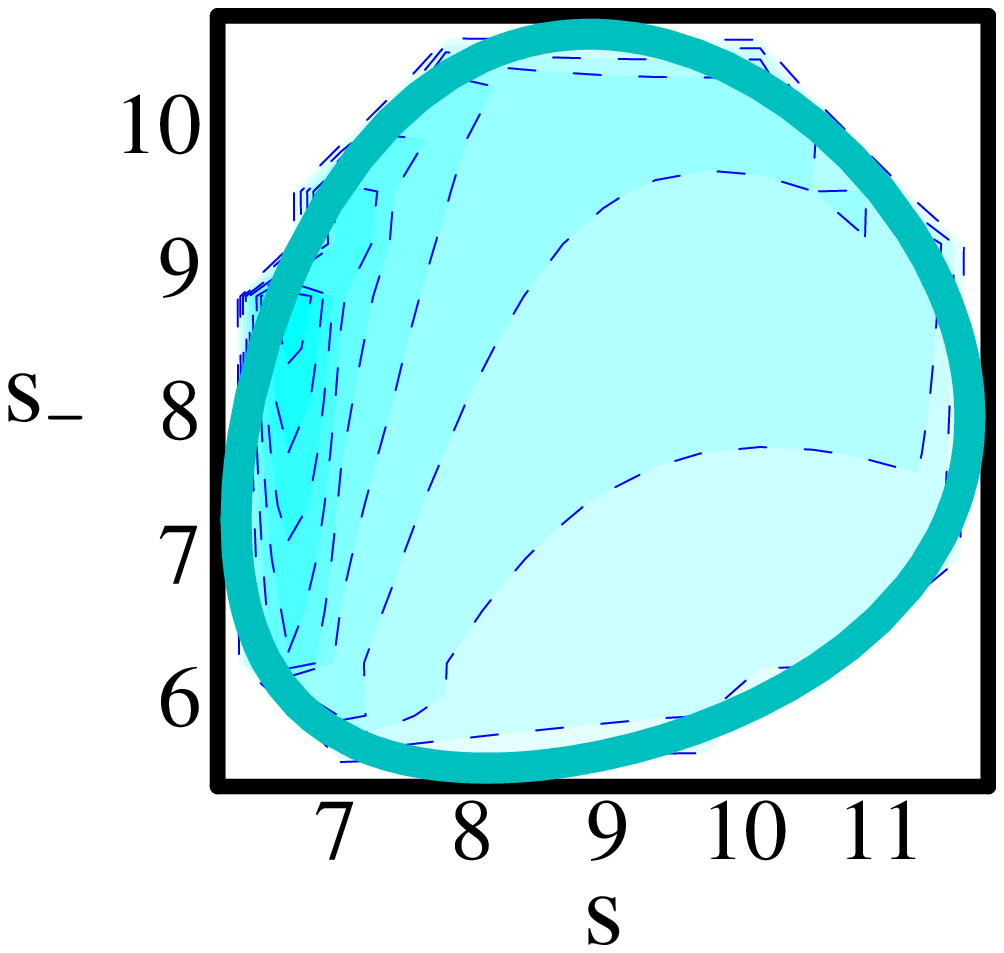, width=4cm}} 
\end{center}
\caption{\baselineskip 15pt 
Differential decay width (left) and Dalitz plot (right) of the transitions
$B^0 \to D^{-} D^{0} K^+$ (up) and  $B^0 \to D^{-} D^{*0} K^+$ (down).
Units are as in fig.\ref{fig:dalitz1}. }
\label{fig:dalitz3}
\vspace*{1.0cm}
\end{figure}

The prominent role of the intermediate states $D_{s0}$ and  $D_{s1}^*$
makes the processes (\ref{channel1}) and  (\ref{channel2}) particularly
promising for the analysis of broad orbital excitations of the $c \bar s$
meson system. It is worth noticing, however, that also other 
three-body $B$ decays can be well suited for such an investigation. Examples
are $B^0 \to D^{-} D^{0} K^+$ and $B^0 \to D^{-} D^{*0} K^+$,
for which no experimental results are available. Since the matrix
element of $B^0 \to D^{-}$ can be related to $B^0 \to D^{*-}$ in the heavy
quark limit, it is possible to determine, in the scheme described in the
previous section, the properties  of the channels $B^0 \to D^{-} D^{(*)0} K^+$.
One predicts:
\bea
{{\cal B}(B^0 \to D^{-}  D^0 K^+) \over 
{\cal B}(B^0 \to D^{*-}  D^0 K^+)}&=&2.11 \nonumber \\ 
{{\cal B}(B^0 \to D^{-}  D^{*0} K^+) \over 
{\cal B}(B^0 \to D^{*-}  D^{*0} K^+)}&=&0.27 
\eea
which imply, considering the experimental data
 in (\ref{eq:res1})-(\ref{eq:res2}):
\bea
{\cal B}(B^0 \to D^{-}  D^0 K^+)&=&(6.3 \pm 1.8) \,\, 10^{-3} \nonumber \\ 
{\cal B}(B^0 \to D^{-}  D^{*0} K^+)&=&(1.8 \pm 0.7) \,\, 10^{-3} \,\,\, ,
\eea
in a range accessible to current experiments. The expected decay 
distributions and the Dalitz plots, depicted in fig.\ref{fig:dalitz3}, 
are similar to those of the modes (\ref{channel1}) and 
(\ref{channel2}), with features that it will be interesting 
to experimentally investigate.
%
%

We conclude this section with a comment on the two main theoretical 
uncertainties in our approach, the use of the heavy quark limit both for
beauty and charm quarks, and the factorization assumed for the 
nonleptonic matrix elements. The two uncertainties are correlated,
and a quantitative assessment of their role is not  a trivial task.
If we consider them separately, we can presume that several 
$\displaystyle{1\over m_Q}$ corrections are compensated in the ratios 
used as the basis of our analysis. As for factorization, the matrix elements
governing the decays considered in this note
 are different from the matrix elements
for which factorization has been proved in the infinite $b$ mass limit
\cite{bbns}. Nevertheless, the study of various processes of the 
type considered here, i.e. color allowed B transitions to charm mesons, 
shows that factorization reproduces the available data within 
their current errors \cite{rosner}. Our analysis can be considered
as a further test of factorization;  experimental measurements will be helpful
in shedding light on the size and the type of possible deviations.

\section{Conclusions}

The analysis of the decays 
$B^0 \to D^{*-} D^{0} K^+$ and $B^0 \to D^{*-} D^{*0} K^+$ shows that they
mainly proceed 
through positive parity intermediate states, $D_{s0}$ and  $D^*_{s1}$.
Other contributions are less significant, so that such three-body $B^0$
transitions  appear to be  well suited for studying the features 
of the low-lying orbital excitations of the $(\bar s c)$ meson system. 
Currently available experimental data allow us to constrain the strong coupling
between such orbital excitations, the negative parity charmed mesons 
and the Kaon, in the region $|h|=0.6\pm 0.2$, close to the expected values. 
An improvement  in the accuracy of the measurements would further constrain
this parameter. On the other hand, the
coupling $g$ is found in a range which extends from zero to 
the CLEO measurement (\ref{gexp}).

The Dalitz plots relative to
$B^0 \to D^{*-} D^{0} K^+$ and $B^0 \to D^{*-} D^{*0} K^+$  are expected 
to display peculiar features, namely the main dependence on the 
invariant $D^{(*)0}K^+$ mass,
that can be experimentally tested. The investigations of 
other modes, $B^0 \to D^{-} D^{0} K^+$ and $B^0 \to D^{-} D^{*0} K^+$, 
can also provide us with information about positive parity charm states; 
their expected decay rates are accessible to current experiments.

\end{document}